\documentclass[
    ,final            
  ,numberedheadings 
  ]
  {aipproc}

\layoutstyle{6x9}

\begin{document}

\title{On the arcmin structure of the X-ray Universe}


\author{J. Ebrero}{
  address={SRON - Netherlands Institute for Space Research, Sorbonnelaan 2, 3584 CA, Utrecht, The Netherlands}
  ,altaddress={Department of Physics and Astronomy, University of Leicester, University Road, LE1 7RH, Leicester, UK}
}

\author{S. Mateos}{
  address={Department of Physics and Astronomy, University of Leicester, University Road, LE1 7RH, Leicester, UK}
}

\author{G. C. Stewart}{
  address={Department of Physics and Astronomy, University of Leicester, University Road, LE1 7RH, Leicester, UK}
}

\author{F. J. Carrera}{
  address={Instituto de F\'isica de Cantabria (CSIC-UC), Avenida de los Castros, 39005, Santander, Spain}
}

\begin{abstract}
We present the angular correlation function of the X-ray population of 1063 XMM-Newton observations at high Galactic latitudes, comprising up to $\sim$30000 sources over a sky area of $\sim$125 sq. degrees in the energy bands: soft (0.5-2 keV) and hard (2-10 keV).
This is the largest sample of serendipitous X-ray sources ever used for clustering analysis purposes to date and the results have been determined with unprecedented accuracy. We detect significant clustering signals in the soft and hard bands ($\sim$10$\sigma$ and $\sim$5$\sigma$, respectively). 
We deproject the angular correlation function via Limber's equation and calculate the typical spatial lengths. We infer that AGN at redshifts $\sim$1 are embedded in dark matter halos with typical masses of $\log M \sim 12.6 h^{-1}$~$M_{\odot}$ and lifetimes in the range $\sim3-5 \times 10^8$~years, which indicates that AGN activity is a transient phase in the life of galaxies.
\end{abstract}

\maketitle


\section{Introduction}

Active galactic nuclei (AGN) are the brightest persistent extragalactic sources known, with their X-ray emission the most common feature among them. Thanks to their large bolometric output, AGN can be detected through cosmological distances, which makes them essential tracers of galaxy formation and evolution, as well as the large-scale structure of the Universe. Clustering studies of AGN at redshift $\sim$1, when strong structure formation processes took place, are key tools for understanding the underlying mass distribution and evolution of cosmic structures.

\section{The X-ray data}

In this work we use the sample presented in \citep{Mat08}. The selected XMM-Newton/EPIC-pn observations fulfilled the following criteria: 1. High galactic latitude fields ($\left|b\right|>20^\circ$) in order to minimize the contamination from Galactic sources; 2. Fields with at least 5~ks of clean exposure time; and 3. Fields free of bright and/or extended X-ray sources.

%
If there were observations carried out at the same sky position, the overlapping area from the observation with the shortest clean exposure time was removed. The resulting sample comprised 1129 observations. For the purposes of this work we have also removed the observations belonging to the Virgo Cluster, M31, M33, Large Magellanic Cloud and Small Magellanic Cloud fields, ending up with a final sample of 1063 observations. The overall sky coverage of the sample is 125.52~deg$^2$ comprising 31288 and 9188 sources in the soft (0.5-2~keV) and hard (2-10~keV) bands, respectively.

\section{The angular correlation function}

The two-point angular correlation function $w(\theta)$ determines the joint probability of finding two objects in two small angular regions $\delta\Omega_1$ and $\delta\Omega_2$ separated by an angular distance $\theta$ with respect to that of a random distribution (\citep{Pee80}).

To calculate the angular correlation function we have used the estimator proposed by \citep{Landy93} $w(\theta_i)=\frac{DD-2DR+RR}{RR}$, where $DD$, $DR$ and $RR$ are the normalised number of pairs of sources in the $i$-th angular bin for the Data-Data, Data-Random and Random-Random samples, respectively. To produce the random source sample against which we have compared the real source sample searching for overdensities at different angular distances, we have tried to mimic as closely as possible the real distribution of the detection sensitivity of the survey. The method employed here is extensively described in \citep{Carrera07} and \citep{Ebr09}. The errors in different angular bins are not independent from one another. To estimate the errors we have followed the covariance matrix method described in \citep{Miy07}.

The angular correlation function can be described by a power-law model in the form $w_{model}(\theta)=\left(\frac{\theta}{\theta_0}\right)^{1-\gamma}$, where $1-\gamma$ is the slope and $\theta_0$ is the angular correlation length. We have fitted the data using a $\chi^2$ technique.

\section{Results}

In the soft band (0.5-2~keV) we detect a high-significance ($\sim$10$\sigma$) clustering signal with a correlation length of $\theta_0=22.9\pm2.0$~arcsec and a slope of $\gamma-1=1.12\pm0.04$ after correcting for the integral constraint. If we ignore that correction in our fits, the correlation lengths are comparable within the error bars but the power-law becomes significantly steeper ($\gamma-1=1.29\pm0.04$).

In the hard band (2-10~keV) the power-law becomes steeper ($\gamma-1=1.33$) and the clustering is stronger ($\theta_0=29.2_{-5.7}^{+5.1}$~arcsec) although marginally consistent with the results in the soft band within the 1$\sigma$ error bars. The clustering detection is still very significant ($\sim$5$\sigma$). The sources detected in the hard band are less biased against absorption, and if the unified model of AGN is correct (the obscuration of the nucleus is due to orientation effects only) one might not expect significant differences in the clustering properties of obscured and unobscured sources. However, accurate angular clustering measurements in this band have been difficult because of the limitations caused by the small-number statistics (e.g. \citep{Gandhi06}, \citep{Carrera07}, \citep{Ueda08}).


\begin{figure}
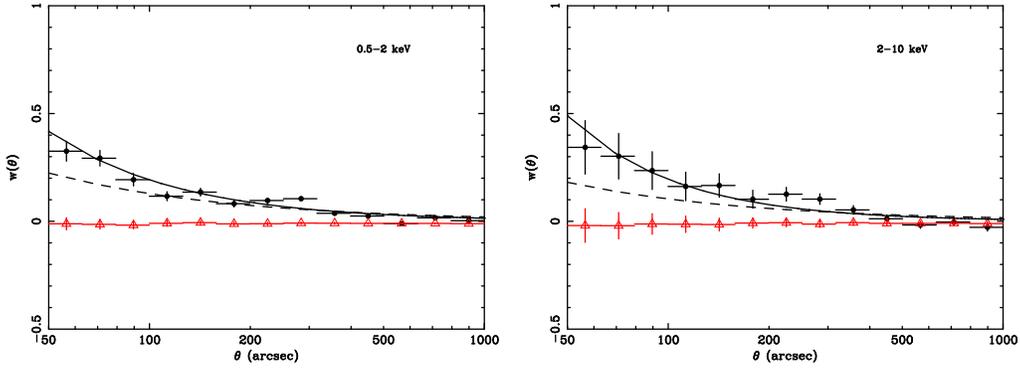

\hbox{
  \includegraphics[height=.3\textheight,angle=-90]{wtsoft_mcov.ps}
}
\hbox{
  \includegraphics[height=.3\textheight,angle=-90]{wthard_mcov.ps}
}
  \caption{Angular correlation function in the 0.5-2~keV (left panel) and 2-10~keV (right panel) bands. Solid dots are the observed data. Triangles represent the averaged $w(\theta)$ for a random dataset. Overplotted is the best-fit $\chi^2$ with and without fixed slope (dashed and solid lines, respectively).}
\label{wtheta}
\end{figure}

\section{Inversion of Limber's equation}

The two-dimensional angular correlation function is a projection in the sky of the real three-dimensional spatial correlation function $\xi(r)$ along the line of sight, where $r$ is the physical separation between sources (typically in units of $h^{-1}$~Mpc). The angular amplitude $\theta_0$ can be related to the spatial amplitude $r_0$ by inverting Limber's integral equation (\citep{Peebles93}). For this, we assumed a redshift selection function for our sample derived from the best-fit luminosity-dependent density evolution (LDDE) model of the X-ray luminosity function of \citep{Ebr08} for all bands under study. Hence, we found values of $r_0=$$12.25\pm0.12$ and $9.9\pm2.4$~ $h^{-1}$~Mpc in the soft and hard bands, respectively, for a clustering model constant in comoving coordinates, while for clustering constant in physical coordinates we obtained $r_0=$$6.54\pm0.06$ and $5.7\pm1.4$~$h^{-1}$~Mpc, respectively.

\section{Dark matter haloes and the lifetime of AGN}

The spatial clustering values derived above can be used to estimate the mass of the dark matter haloes (DMH) in which these sources are embedded. A commonly used quantity for such an analysis is the bias parameter, that is usually defined as $b^2(z)=\frac{\xi_{AGN}(8,z)}{\xi_{DMH}(8,z)}$, where $\xi_{AGN}(8,z)$ and $\xi_{DMH}(8,z)$ are the spatial correlation functions of AGN and DMH evaluated at 8 $h^{-1}$~Mpc, respectively. The former value has been calculated in this work whereas the latter can be estimaded using the expressions in \citep{Pee80}. In a $\Lambda$-CDM cosmology, the bias is function of both the mass of DMH and redshift (\citep{MW01}). From this we derive an average mass of $\langle \log M_{DMH}\rangle \simeq 12.60\pm0.34$~$h^{-1}$~M$_\odot$. This result is fully in agreement with that of \citep{Gilli09} who found $\log M_{DMH}=12.4-12.8$~$h^{-1}$~M$_\odot$ in the COSMOS survey, although it is worth noticing that our results come from a mostly unidentified X-ray sample. Similarly, if we compare our results with the clustering study of \citep{Coil09} in the AEGIS field we see that our mean DMH mass is intermediate between that of blue and red galaxies, which is in agreement with the location of X-ray selected AGN in the color-magnitude diagram. We can estimate the lifetime of AGN using the mean DMH masses calculated above and making some simple assumptions. We have followed the method proposed by \citep{MW01}, assuming that we are sampling the most massive DMH at a given redshift $z$ and that each DMH hosts an active AGN at any given time. Hence we have $t_{AGN}(z)=t_U(z)\frac{\Phi(z)}{\Phi_{DMH}(z)}$, where $t_U(z)$ is the Hubble time at redshift $z$, $\Phi(z)$ is the comoving density of AGN for which we used the predictions of the luminosity function of \citep{Ebr08}, and $\Phi_{DMH}(z)$ is the comoving density of DMH that can be estimated following the Press-Schechter approximation. At $z \sim 1$ and for haloes larger than $\log M_{min} = 12.6$~$h^{-1}$~M$_\odot$, we get $\Phi_{DMH} \simeq 2 \times 10^{-3}$~$h^3$~Mpc$^{-3}$. This yielded to an AGN duty cycle in the range $t_{AGN}/t_U = 0.054-0.078$. In our cosmological framework, the Hubble time at $z=1$ is $\sim$5.8~Gyr. The estimated lifetime of AGN is hence in the range $t_{AGN}=3.1-4.5 \times 10^8$~yr.

\section{Conclusions}
\label{conclusions}

We have studied the angular correlation function of a large sample of serendipitous X-ray sources from 1063 XMM-Newton observations at high Galactic latitudes in the 0.5-2~keV (soft), and 2-10~keV (hard) energy bands, respectively. Our sample comprises 31288 and 9188 sources in the soft and hard bands, respectively, covering $\sim$125.5~deg$^{2}$ in the sky, thus being the largest sample ever used in clustering investigations.

We found significant positive angular clustering signal in the soft ($\sim$10$\sigma$) and hard ($\sim$5$\sigma$) bands. The result in the hard band clears up the debate on whether X-ray sources detected in this band cluster or not, since a number of past works had reported different inconclusive results ranging from a few $\sigma$ detections to no detection at all.

We inverted Limber's equation, assuming a given redshift distribution for our sources, in order to estimate typical spatial correlation lengths. We used these values to calculate the $rms$ fluctuations of the AGN distributions within a sphere of radius 8 $h^{-1}$~Mpc, and compared them with that of the underlying mass distribution from the linear theory in order to estimate the bias parameter of our X-ray sources. The bias depends on the mass of the dark matter haloes (DMH) that host the AGN population. From the computed bias values we have estimated a typical DMH mass of $\langle \log M_{DMH}\rangle \simeq 12.60\pm0.34$~$h^{-1}$~M$_\odot$.

The typical AGN lifetime derived from the Press-Schechter approximation at redshift $z \sim 1$ lies in the range in the range $t_{AGN}=3.1-4.5 \times 10^8$~yr. This interval is significantly shorter than the time span between that redshift and the present thus suggesting the existence of many AGN generations, and that a significant fraction of galaxies may switch from a quiescent phase to AGN activity, and vice versa, several times throughout their lifes.





\bibliographystyle{aipproc}   



\end{document}